\documentclass[11pt]{article}

\usepackage{graphicx,amssymb,amsmath}
\usepackage{latexsym}
\usepackage{color}
\usepackage{fullpage}

\long\def\ignore#1{}

\begin{document}

\title{An $O({\log n\over \log\log n})$ Upper Bound on the Price of Stability for Undirected
Shapley Network Design Games}
\author{
Jian Li}

\maketitle

\begin{center}
University of Maryland, College Park
\end{center}

\begin{abstract}
In this paper, we consider the Shapley network design game on
undirected networks.
In this game, we have an edge weighted undirected network $G(V,E)$ and $n$ selfish players
where player $i$ wants to choose a path from source vertex $s_i$ to
destination vertex $t_i$. The cost of each edge is equally split
among players who pass it. The price of stability is defined as
the ratio of the cost of the best Nash equilibrium to that of the optimal solution.
We present an $O(\log n/\log\log n)$
upper bound on price of stability for the single sink case, i.e, $t_i=t$ for all $i$.
\end{abstract}


\section{Introduction}
We consider the Shapley network design game, which is also called network design games with
fair cost allocation, introduced in \cite{FOCS04:pos}.
In this game, we are given a network and $n$ selfish players.
where player $i$ wants to go from source vertex $s_i$ to
destination vertex $t_i$. The cost of each edge is shared
in a fair manner among players who pass it.
We are interested in stable status of the network
where no player has the incentive to deviate from its current strategy,
which can be modeled by Nash equilibria.
The price of stability, defined as the ratio of the cost of
the best Nash equilibrium and that of an optimal solution, is used to measure
the inefficiency of Nash equilibria.
We imagine a network where the traffic will be initially
designed by a central network coordinator.
However, the coordinator is unable prevent the network users from
selfishly deviating from the designated paths.
Therefore, in this scenario, the best Nash equilibrium is an obvious
solution to propose.
In this sense, we can think the price of stability as the degree of degradation of the solution quality
for the outcome being stable.

The price of stability was first studied
in Schulzan and Moses \cite{SODA03:SM} and was so-called in Anshelevich
et al. \cite{FOCS04:pos} where the Shapley network design game was also first explored.
They showed that a pure-strategy Nash equilibrium always exists and
the price of stability of this game is at most the $n$th harmonic number
$H(n)$ and also provide an example showing that this
upper bound is the best possible in directed networks.
For undirected networks, Anshelevich et al. \cite{FOCS04:pos} presented
a tight bound on price of stability of $4/3$ for single source and two players case.
However, whether there is a tighter bound for arbitrarily many players in
undirect networks was left as an open question.
Fiat et al. \cite{icalp06:pos} improved the upper bound to $O(\log\log n)$
for a special case where each node of the network has a player and they are required to connect
to a common destination. Chen and Roughgarden \cite{SPAA06:ho_tim} considered the weighted version
of the game where each player has a weight and the cost of an edge is shared among
the players who pass it in proportion to their weights.
As opposed to the ordinary Nash equilibrium considered before,
Albers \cite{SODA08:susan} investigated the situation where coordination among players is
allowed and showed nearly matching upper and lower bounds on the price of stability with respect to
the notion of {\em strong Nash equilibrium}.

\noindent\textbf{Our results :} We prove that for undirected graphs with a distinguished
destination to which all players must connect, the price of stability of the Shapley network design game
is $O({\log n\over \log\log n})$ where $n$ is the number of players.




\section{Preliminaries}
\label{sec:prel}

We first introduce notations and formally state the problem.
We are given a undirected network $G(V,E)$ and $n$ selfish players.
Player $i$ has to choose a path from source vertex $s_i$
to destination vertex $t_i$.
Let $\mathcal{P}_i$ denote the set of simple $s_i-t_i$ paths.
The cost of an edge $e$, $c(e)$, is shared equally by all players who pass $e$.
An {\em outcome} of the game is specified by a set of $n$ path, each chosen by one player.
For an outcome $(P_1,P_2,\ldots,P_n)$ for $P_i\in \mathcal{P}_i$
,the cost assigned to player $i$ is $c_i(P_1,P_2,\ldots,P_n)=\sum_{e\in P_i}{c_e\over f_e}$ where
$f_e$ is the number of paths that include $e$.
We define the cost of the outcome as
$$
c(P_1,P_2,\ldots,P_n)=\sum_i c_i(P_1,P_2,\ldots,P_n)=\sum_{e\in \cup_i P_i}c_e.
$$

Let $P_{-i}$ denote the vector of paths chosen by the players other than $i$
An outcome $(P_1,P_2,\ldots,P_n)$ is a \emph{Nash equilibrium} if for every player $i$
, $c_i(P_i,P_{-i})=\min_{\tilde{P}_i\in \mathcal{P}_i}c_i(\tilde{P}_i,P_{-i})$.

The \emph{price of stability} is defined as the ratio of the cost of the best Nash
equilibrium of the game to that of an optimal solution.
We note that the optimal solution is the min-cost steiner forest satisfying all
connectivity requirement $(s_i,t_i)$s.

We consider the following potential function, also used in \cite{FOCS04:pos},
that maps every outcome into a numeric value.

\begin{equation}
\label{eq:nd}
\Phi(P_1,\ldots,P_k)=\sum_{e\in E}\sum_{i=1}^{f_e}{c_e\over
i}=\sum_{e\in E}c_e\cdot H(f_e)
\end{equation}
where $f_e$ denotes the number of paths $P_i$ that include edge $e$ and
$H(n)=1+{1\over 2}+{1\over 3}+\ldots+{1\over n}$ is the $n$'th Harmonic number.

The most important property of the potential function is that
if a single player $i$ changes its strategy then the difference between the potential
of the new state and that of the original state is exactly the change in the cost of player $i$ \cite{FOCS04:pos}.


In a finite game, \emph{better-response dynamics} is the following process: If the current outcome
is not a Nash equilibrium, there exists a player who can decrease its cost by switching its strategy.
The player updates its strategy to an arbitrary superior one, and repeat until a Nash equilibrium is reached.
While better response dynamics needs not terminate in general, it
must terminate in finite steps in Shapley network design games since the potential $\Phi$
strictly decreases during the process and no outcome appears twice in a finite game.

\section{An $O({\log n\over\log\log n})$ Upper Bound for the Single Sink Case}
\label{sec:main}

We assume the network is connected and all players share the same destination $t$.
It is easy to see an optimal solution is a steiner tree with terminals $\{s_i\}_{i=1,\ldots,n}\cup\{t\}$.
Suppose the outcome $NASH=(P^N_1,\ldots,P^N_n)$ is a Nash equilibrium
which is obtained by better-response dynamics from an optimal solution $OPT=(P^O_1,\ldots,P^O_n)$.
We can assume without loss of generality that $NASH$ is also a tree.
The property of the the potential function ensures that $\Phi(NASH)\leq \Phi(OPT)$.
We denote paths of $NASH$ and that of $OPT$ by $\{P^N_i\}_{i=1,\ldots,n}$ and $\{P^O_i\}_{i=\{1,\ldots,n\}}$,
respectively. We also denote the trees of $Nash$ by $T^N=\cup_i P^N_i$ and that of $OPT$ by $T^O=\cup_i P^O_i$.
Let $|NASH|$ and $|OPT|$ be their costs respectively.

Let $f_e^N$ denote the number of paths that include edge $e$ in $NASH$.
Let $f^N(i)=\sum_{e: f_e^N=i}c_e$ and $g^{N}(j)=\sum_{e:f_e^N\geq j}c_e=\sum_{i\geq j}f^N(i)$.
It is easy to see $|NASH|=\sum_i f^N(i)=g^N(1)$.

For ease of discussion, we create a dummy player $0$ residing in $s_0=t$.
We can see this player has no influence on either $NASH$ or $OPT$.
First we consider the tree $T_O=\cup_i P^O_i$. Doubling all edges in $T_O$ forms a Eulerian tour.
Traversing this tour gives a sequence $S$ of vertices in $T_O$.  Suppose $\phi$
is a permutation of $\{s_i\}_{i=0,\ldots,n}$ according to their first appearance in $S$.
It is easy to see $\sum_{i=0}^n d(\phi(i),\phi(i+1\textrm{ mod }n+1))\leq 2|T_O|=2|OPT|$ where $d(u,v)$ is the length
of the shortest path between $u$ and $v$.

For any two players $i$ and $j$, let
$LCA(i,j)$ be the least common ancestor of $s_i$ and $s_j$ in tree $T^N$(take $t$ as the root).
We let $P^j_i$ be the subpath of $P^N_i$ starting from $s_i$
and ending at $LCA(i,j)$.
From the definition of Nash equilibrium, we know the cost of player $i$ in $NASH$
is less than that of first reaching $s_j$ and then following the path $P^N_j$ to $t$.
Thus, we have the following.
$$
\sum_{e\in P^j_i}{c_e\over f^N_e} \leq d(s_i,s_j)+ \sum_{e\in P^i_j}{c_e\over f^N_e+1}.
$$
Similarly, we have
$$
\sum_{e\in P^i_j}{c_e\over f^N_e} \leq d(s_i,s_j)+ \sum_{e\in P^j_i}{c_e\over f^N_e+1}.
$$
Adding them together, We get
$$
\sum_{e\in P^j_i}{c_e\over f^N_e(f^N_e+1)}+\sum_{e\in P^i_j}{c_e\over f^N_e(f^N_e+1)}\leq 2d(s_i,s_j).
$$
We denote the left hand side of last equality by $A(i,j)$. We have
$\sum_{i=0}^n A(\phi(i),\phi(i+1\textrm{ mod }n+1))\leq 2\sum_{i=0}^n d(v_i,v_{i+1\textrm{ mod }n+1}) \leq 4|OPT|$.

Now we prove
\begin{equation}
\label{eq:na}
\sum_{i=0}^n A(\phi(i),\phi(i+1\textrm{ mod }n+1))\geq 2\sum_{e\in T_N} {c_e\over f^N_e(f^N_e+1)}
=2\sum_{i}{1\over i(i+1)} f^N(i)
\end{equation}

Actually, we only need to prove every $e\in T_N$ appears in
some $P_{\phi(i)}^{\phi(i+1\textrm{ mod }n+1)}$ or $P_{\phi(i+1\textrm{ mod }n+1)}^{\phi(i)}$ for $0\leq i \leq n$.
It is easy to see $P_i^j\cup P^j_i$ is the unique path from $s_i$ to $s_j$ in $T_N$.
For any $e\in T_N$, let $T^1_{N,e}$ and $T^2_{N,e}$ be two trees obtained by deleting $e$ from $T_N$.
It is easy to see $T^i_{N,e}\cap \{s_0,\ldots,s_n\}\ne \emptyset$ for $i=1,2$ since each leaf of $T_N$
contains at least one player.
So, there exists some $i$ such that $\phi(i)\in T^1_e$ and $\phi(i+1\textrm{ mod }n+1)\in T^2_e$
and $e$ must lie in the unique path from $S_{\phi(i)}$ to $S_{\phi(i+1\textrm{ mod }n+1)}$.

We let $n_{1/2}=\max\{i| g^N(i)\geq {1\over 2}\cdot |NASH|\}$.
We can see the following.
$$
\begin{array}{ll}
\Phi(NASH)& =\sum_i f^N(i)H(i)\geq \sum_{i\geq n_{1/2}}f^N(i)H(i)\geq H(n_{1/2})\sum_{i\geq n_{1/2}}f^N(i) \\
          & =H(n_{1/2})g^N(n_{1/2}) \geq {1\over 2}H(n_{1/2})|NASH|.
\end{array}
$$
Since $\Phi(NASH)\leq \Phi(OPT)\leq H(n)|OPT|$, we have
\begin{equation}
\label{eq:eq1}
|NASH|\leq {2H(n)\over H(n_{1/2})}\cdot|OPT|.
\end{equation}
From (\ref{eq:na}), we can get
$$
\begin{array}{ll}
2|OPT|    & \geq \sum_{i}{1\over i(i+1)} f^N(i) \geq \sum_{i< n_{1/2}}{1\over i(i+1)} f^N(i)\geq {1\over n_{1/2}(n_{1/2}+1)} \sum_{i\leq n_{1/2}}f^N(i)\ \\
          & \geq {1\over 2n_{1/2}(n_{1/2}+1)} \sum_i f^N(i) ={1\over 2n_{1/2}(n_{1/2}+1)}|NASH|.
\end{array}
$$
Thus, we have
\begin{equation}
\label{eq:eq2}
|NASH|\leq 4n_{1/2}(n_{1/2}+1)\cdot |OPT| .
\end{equation}

Combining inequalities (\ref{eq:eq1}) and (\ref{eq:eq2}), we have
$|NASH|\leq \min\{ {2H(n)\over H(n_{1/2})},4n_{1/2}(n_{1/2}+1)\}\cdot |OPT|$
for any $n_{1/2}$. The right hand side takes maximum
value $O({\log n\over \log\log n})\cdot |OPT|$ by choosing $n_{1/2}=O(\sqrt{{\log n\over \log\log n}})$.
Therefore, we have proved ${|NASH|\over |OPT|}\leq O({\log n\over \log\log n})$.

\end{document}